\documentclass[twocolumn]{aastex631}

\usepackage{ulem}
\usepackage{amsmath}

\shorttitle{JWST Observations of JAGB Stars}
\shortauthors{Lee et al.}
\graphicspath{{./}{figures/}}

\begin{document}

\title{First JWST Observations of JAGB Stars in the SN Ia Host Galaxies:\\ NGC 7250, NGC 4536, NGC 3972}

\author{Abigail~J.~Lee}\affil{Department of Astronomy \& Astrophysics, University of Chicago, 5640 South Ellis Avenue, Chicago, IL 60637}\affiliation{Kavli Institute for Cosmological Physics, University of Chicago,  5640 South Ellis Avenue, Chicago, IL 60637}

\author{Wendy~L.~Freedman}\affil{Department of Astronomy \& Astrophysics, University of Chicago, 5640 South Ellis Avenue, Chicago, IL 60637}\affiliation{Kavli Institute for Cosmological Physics, University of Chicago,  5640 South Ellis Avenue, Chicago, IL 60637}

\author{In~Sung~Jang}\affil{Department of Astronomy \& Astrophysics, University of Chicago, 5640 South Ellis Avenue, Chicago, IL 60637}\affiliation{Kavli Institute for Cosmological Physics, University of Chicago,  5640 South Ellis Avenue, Chicago, IL 60637}

\author{Barry~F.~Madore}\affil{Observatories of the Carnegie Institution for Science 813 Santa Barbara St., Pasadena, CA~91101}\affil{Department of Astronomy \& Astrophysics, University of Chicago, 5640 South Ellis Avenue, Chicago, IL 60637}

\author{Kayla~A.~Owens}\affil{Department of Astronomy \& Astrophysics, University of Chicago, 5640 South Ellis Avenue, Chicago, IL 60637}\affiliation{Kavli Institute for Cosmological Physics, University of Chicago,  5640 South Ellis Avenue, Chicago, IL 60637}

\correspondingauthor{Abigail J. Lee}\email{abbyl@uchicago.edu}

\begin{abstract}
The J-region Asymptotic Giant Branch (JAGB) method is a standard candle that leverages the constant luminosities of color-selected, carbon-rich AGB stars, measured in the near infrared at 1.2 microns.
The Chicago-Carnegie Hubble Program (CCHP) has obtained JWST imaging of the SN~Ia host galaxies NGC\,7250, NGC 4536, and NGC 3972.
With these observations, the JAGB method can be studied for the first time using JWST.  \cite{2022ApJ...933..201L} demonstrated the JAGB magnitude is optimally measured in the outer disks of galaxies, because in the inner regions the JAGB magnitude can vary significantly due to a confluence of reddening, blending, and crowding effects. However, determining where the `outer disk' lies can be subjective. Therefore, we introduce a novel method for systematically selecting the outer disk. In a given galaxy, the JAGB magnitude is first separately measured in concentric regions, and the `outer disk' is then defined as the first radial bin where the JAGB magnitude stabilizes to a few hundredths of a magnitude. After successfully employing this method in our JWST galaxy sample, we find the JAGB stars are well-segregated from other stellar populations in color-magnitude space, and have observed dispersions about their individual $F115W$ modes of $\sigma_{N7250}=0.32$~mag, $\sigma_{N4536}=0.34$~mag, and $\sigma_{N3972}=0.35$~mag. These measured dispersions are similar to the scatter measured for the JAGB stars in the LMC using 2MASS data ($\sigma=0.33$~mag, \citealt{2001ApJ...548..712W}).
In conclusion, the JAGB stars as observed with JWST clearly demonstrate their considerable power both as high-precision extragalactic distance indicators and as SN~Ia supernova calibrators. 
\end{abstract}


\section{Introduction}
Determining a consistent and accurate value for the expansion rate of the universe, as parameterized by the Hubble constant ($H_0$), has proven to be an extraordinarily difficult endeavor.  
Measurements of $H_0$ from local distance ladders compared with indirectly inferred measurements from the Cosmic Microwave Background \citep{2020A&A...641A...6P, 2020JCAP...12..047A} currently disagree at the $5\sigma$ level, potentially indicating new physics and theories \citep{ 2021arXiv210301183D}.
However, significant evidence still points to the possibility that systematic uncertainties in the local distance scale may be the reason for this tension. For example, the current two most precise local distance ladder measurements of $H_0$ from the Tip of the Red Giant Branch (TRGB) by the CCHP \citep{2021ApJ...919...16F} and from the Cepheid Leavitt Law by the SH0ES group \citep{2022ApJ...934L...7R} themselves differ by about $2\sigma$. It would seem prudent to resolve this local problem (involving stars) before invoking new physics (involving the foundations of contemporary cosmology).

One means of uncovering  systematic errors in the local (stellar) distance ladder is by comparing the individual distances measured by the Cepheids and TRGB stars with a third equally precise and accurate distance indicator. Pertinently, two independent studies in the recent literature have presented a novel means of measuring distances to galaxies using carbon-rich, asymptotic giant branch stars \citep{2020ApJ...899...66M, 2020ApJ...899...67F, 2020MNRAS.495.2858R}, denominated the J-region Asymptotic Giant Branch (JAGB) method. JAGB stars are a well-defined class of carbon-rich, thermally-pulsating AGB stars whose narrowly-defined, constant luminosity in the near infrared (NIR) has been shown to be constant from galaxy to galaxy. 
Easily identified on the basis of their NIR colors, the mean luminosity of a galaxy's JAGB stars provide us with an excellent and simple, empirical standard candle \citep{2000ApJ...542..804N, 2001ApJ...548..712W}.

Carbon stars' narrow range of luminosities have been predicted by stellar modeling for more than half a century \citep{1973ApJ...185..209I, 1983ARA&A..21..271I}. JAGB stars are also so photometrically distinctive due to their extremely red color, a product of their carbon-enhanced atmospheres. This phenomenon results from the AGB stars' \textit{third dredge-up} event, which transports carbon from the helium-burning shells up to the stellar surface through convective pulses \citep{2003agbs.conf.....H}. The third dredge-up event is only effective for a small range of AGB star masses ($2-5 M_{\odot}$) and thus a small range of luminosities (e.g., \citealt{ 2008A&A...482..883M, 2020ApJ...899...66M}), therefore explaining the JAGB stars' low intrinsic scatter in their NIR luminosities ($\pm0.3$ mag). 

The JAGB method has been shown to have comparable accuracy and precision in measuring distances to nearby galaxies with the Leavitt Law and TRGB \citep{2020ApJ...899...67F, 2020arXiv201204536L, 2021MNRAS.501..933P,2021arXiv210502120Z, 2022ApJ...933..201L, 2023MNRAS.522..195P}, and therefore will be a useful cross-check for both of these distance indicators. The JAGB method also has several clear advantages as a distance indicator: first, JAGB stars are brighter in the near infrared than both the TRGB and 10-day Cepheids. 
Second, JAGB stars are mostly ubiquitous in all galaxy morphologies\footnote{JAGB stars are found in all galaxies with stellar populations between 200 Myr to 1 Gyr. Therefore, they will not be found in old ellipticals and some dwarf galaxies.} and inclinations, unlike the Cepheid Leavitt Law which can only be applied to late-type spiral galaxies with low to moderate inclinations. Third, only one epoch of observing is needed to measure a JAGB distance, unlike Cepheids and Miras for which more than a dozen observations are required to measure their periods, amplitudes, mean magnitudes, and colors. 

In this paper we show results for the first JWST NIRCam \citep{2023PASP..135d8001R, 2023PASP..135b8001R, 2023arXiv230404869G} observations of JAGB stars in the first three galaxies of our JWST Cycle 1 program 1995 (PI: W. Freedman): NGC 7250, NGC 4536, and NGC 3972. The purpose of this program was to measure distances to 10 SN Ia host galaxies with three independent distance indicators, using the same imaging. Each of these methods, the Cepheids, the TRGB method, and the JAGB method, have been shown to be high-precision extragalactic distance indicators. Our research program, the CCHP, will provide a path to $H_0$ based on these three independent methods, aiming to decrease the systematic uncertainties below the 2\% level for each method. This paper in particular marks the beginning of a future measurement of $H_0$ via the JAGB stars.

\section{Data}\label{sec:data}

Data presented here were taken with the JWST's NIRCam. Imaging were simultaneously obtained in the   short-wavelength channel (pixel scale of $0.031\arcsec ~\rm{pixel^{-1}}$) and the long-wavelength channel ($0.063\arcsec ~\rm{pixel^{-1}}$).  The $F115W$ filter (J band equivalent) was chosen for this program as the short-wavelength filter following \cite{2020ApJ...899...66M}, who found that the JAGB stars have a constant mean luminosity in the J band. 
$F444W$ was originally chosen as the long-wavelength filter to have a large $(F115W-F444W)$ color baseline to easily separate O-rich AGB stars and C-rich AGB stars via their colors. The first two galaxies in our program to be imaged, NGC 7250 and NGC 4536, therefore have $F115W$ and $F444W$ data. However, it became clear that the pixel resolution of $F444W$ was poorer than even the HST WFC3/IR camera. We elected to therefore change our LW filter to $F356W$ for the remaining nine galaxies to take advantage of its increased resolution over $F444W$. We found that the $(F115W-F356W)$ color was still excellent at separating the O-rich and C-rich AGB stars in NGC 3972.

Details on the calibration and the reduction of these data are extensively described in a companion paper, Owens et al. (2023, submitted). In short, PSF photometry was extracted from the images using the software package DOLPHOT \citep{2000PASP..112.1383D, 2016ascl.soft08013D} via the the beta testing JWST/NIRCAM module \citep{2023arXiv230104659W}.

The photometry used in this analysis remains preliminary for two reasons: photometry was extracted via the beta-testing DOLPHOT JWST/NIRCAM module (which is still being actively updated; D. Weisz, private communication), and the absolute photometric zeropoints of NIRCam still remain relatively uncertain \citep{2022RNAAS...6..191B}. A JWST cycle 1 proposal aims to more accurately calibrate the NIRCam photometric zeropoints \citep{2022AJ....163..267G}, after which we expect to be able to publicly provide final photometry closer to the requisite $<2\%$~precision needed for extragalactic distance scale measurements of $H_0$.

Nevertheless, even while using this preliminary photometry, the JAGB stars still exhibit low scatter and are clearly delineated from other stellar populations. However, we still cautiously emphasize that the magnitude axes should be taken as arbitrary. We have added the same additional random magnitude offset between $-0.1$ and $0.1$~mag to both our F115W and F444W/F356W photometry. This offset will be removed once our photometric zeropoints are finalized in the next paper in this series. Below, we now describe the data for the three galaxies analyzed in this paper.

\begin{figure*}
\centering
\includegraphics[width=\textwidth]{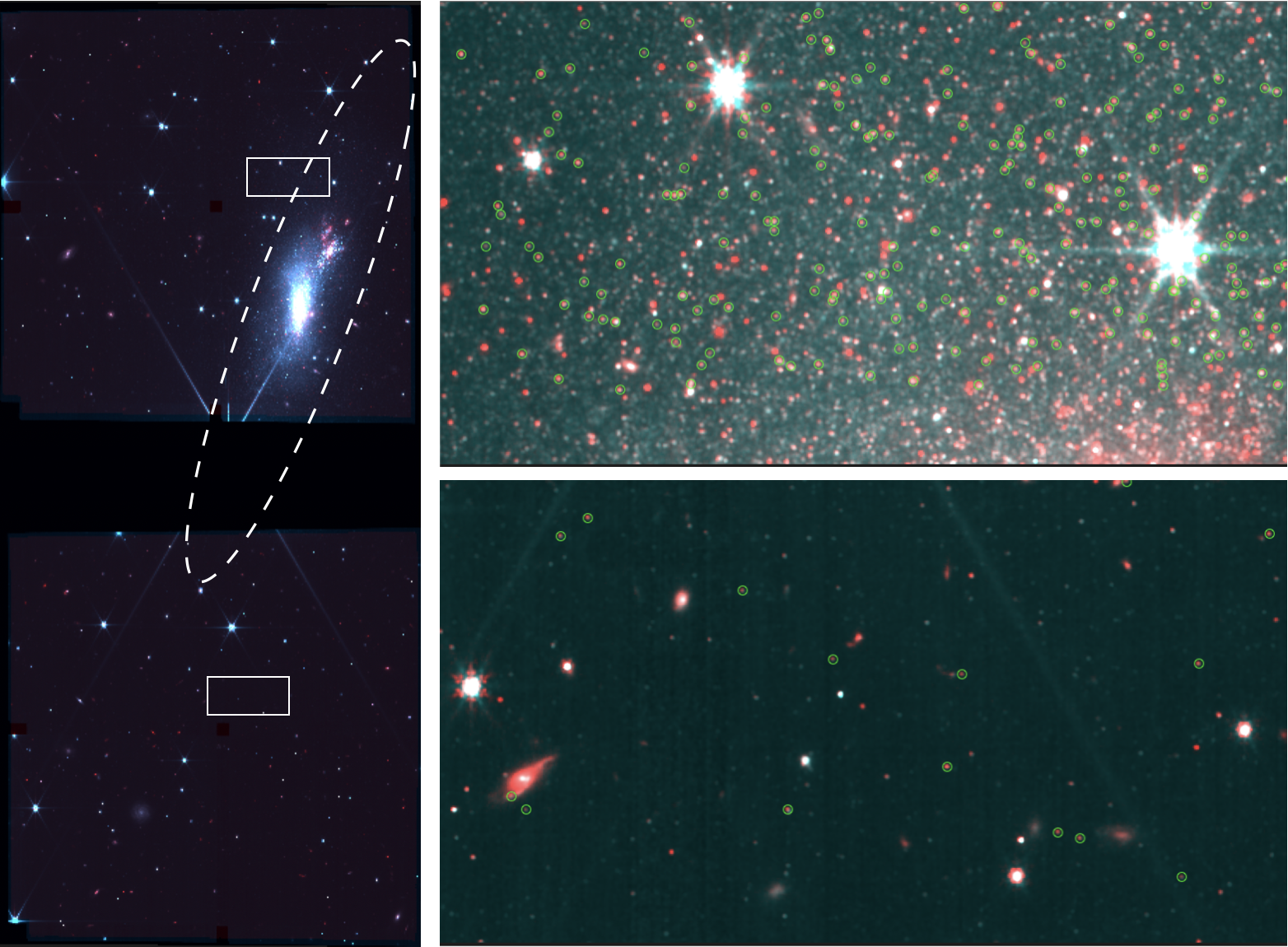}
\caption{Left: a color mosaic of the  $F115W$ and  $F444W$ images for the observed NGC\,7250 field. 
Our analysis excluded the stars in NGC 7250's inner disk, delineated by the dotted white ellipse. Right: zoom-in views of the inner (top) and outer (bottom) rectangles of the left image. The JAGB stars are highlighted within the green circles. North is up and east is to the left in all the panels. }
\label{fig:montage}
\end{figure*}

\subsection{NGC 7250}
The irregular galaxy NGC 7250 (also known as PGC 68535 or UGC 11980) was the first SN Ia host galaxy in this program to be observed by JWST in November 2022. Total exposures of 3769 seconds were simultaneously obtained in the  $F115W$ short-wavelength channel and the $F444W$ long-wavelength channel. A color mosaic image of the observations is shown in Figure \ref{fig:montage}. In 2013, NGC~7250 was host to the type 1a SN 2013dy explosion, therefore uniquely positioning it as a SN Ia calibrator for measuring the Hubble constant. Indeed, \cite{2016ApJ...826...56R, 2022ApJ...934L...7R} used NGC~7250 as a SN Ia calibrator using 21 Cepheids observed with the Hubble Space Telescope (HST) as part of the SH0ES program, measuring a Leavitt law distance modulus of $31.628\pm0.126$~mag (or $21.2\pm1.3$~Mpc). 
Photometry for $\sim514,000$~ sources was extracted from the images; after performing cleaning cuts designed to remove non-stellar sources (e.g., artifacts, cosmic rays, and extended sources) (DOLPHOT type = 1, $\rm{SNR}_{F115W}>3$, $\rm{SNR}_{F444W}>1$, $\rm{sharp}^2 <0.04$, $\rm{crowd}_{F115W}<0.2$, $\sigma_{F115W}<0.02 \times 0.0025 \rm{e}^{F115W-24.05}$), $\sim112,000$ sources remained in the final catalog. In particular, our chosen quality cuts do a reasonable job of eliminating background galaxies, but we plan to more thoroughly investigate the optimal quality cut parameters in our full sample of SN Ia host galaxies. We note, however, that the mode (used to measure the JAGB magnitude) is particularly robust to contamination from background galaxies, which are mostly $\sim 1$ mag fainter than the JAGB stars \citep{2022ApJ...926..153M}.

\subsection{NGC 4536}
NGC 4536 (also known as PGC 41823 or UGC 07732) is a spiral galaxy that was observed with JWST in January 2023, and was host to SN 1981B. Its distance was measured by the CCHP via the TRGB (\citealt{2018ApJ...861..104H}, $M_{F814W}^{TRGB}=-4.05\pm0.04$~mag, $A_{F814W}=0.03$~mag) to be $\mu_0 = 31.04\pm0.06$~mag (or $16.1\pm0.5$~Mpc). Its distance was also measured using the Leavitt law via 40 Cepheids \citep{2016ApJ...826...56R, 2022ApJ...934L...7R} to be $\mu_0=30.838\pm0.051$~mag (or $14.7\pm0.3$~Mpc). Total exposures of 2802 seconds were simultaneously obtained in the $F115W$ short-wavelength channel and the $F444W$ long-wavelength channel.
Photometry for $\sim 566,000$ sources was extracted from the images; after performing cleaning cuts (DOLPHOT type = 1, $\rm{SNR}_{F115W}>3$, $\rm{SNR}_{F444W}>1$, $\rm{sharp}^2 <0.04$, $\rm{crowd}_{F115W}<0.2$), $\sim210,000$ sources remained in the final catalog.  

\subsection{NGC 3972}
NGC 3972 (also known as UGC 6904) is a spiral galaxy that was observed with JWST in April 2023, and was host to SN 2011by. \cite{2016ApJ...826...56R, 2022ApJ...934L...7R} used 52 Cepheids to measure a distance modulus of $\mu_0=31.644\pm0.090$~mag (or $21.3\pm0.9$~Mpc). Total exposures of 3768 seconds were simultaneously obtained in the $F115W$ short-wavelength channel and the $F356W$ long-wavelength channel. Photometry for $\sim 688,000$ sources was extracted from the images; after performing cleaning cuts (DOLPHOT type = 1, $\rm{SNR}_{F115W}>3$, $\rm{SNR}_{F444W}>1$, $\rm{sharp}^2 <0.04$, $\rm{crowd}_{F115W}<0.2$, $\sigma_{F115W}<0.02 \times 0.0025 \rm{e}^{F115W-24.05}$), $\sim78,000$ sources remained in the final catalog.

\section{Measuring the JAGB magnitude}

\begin{figure*}
\gridline{\fig{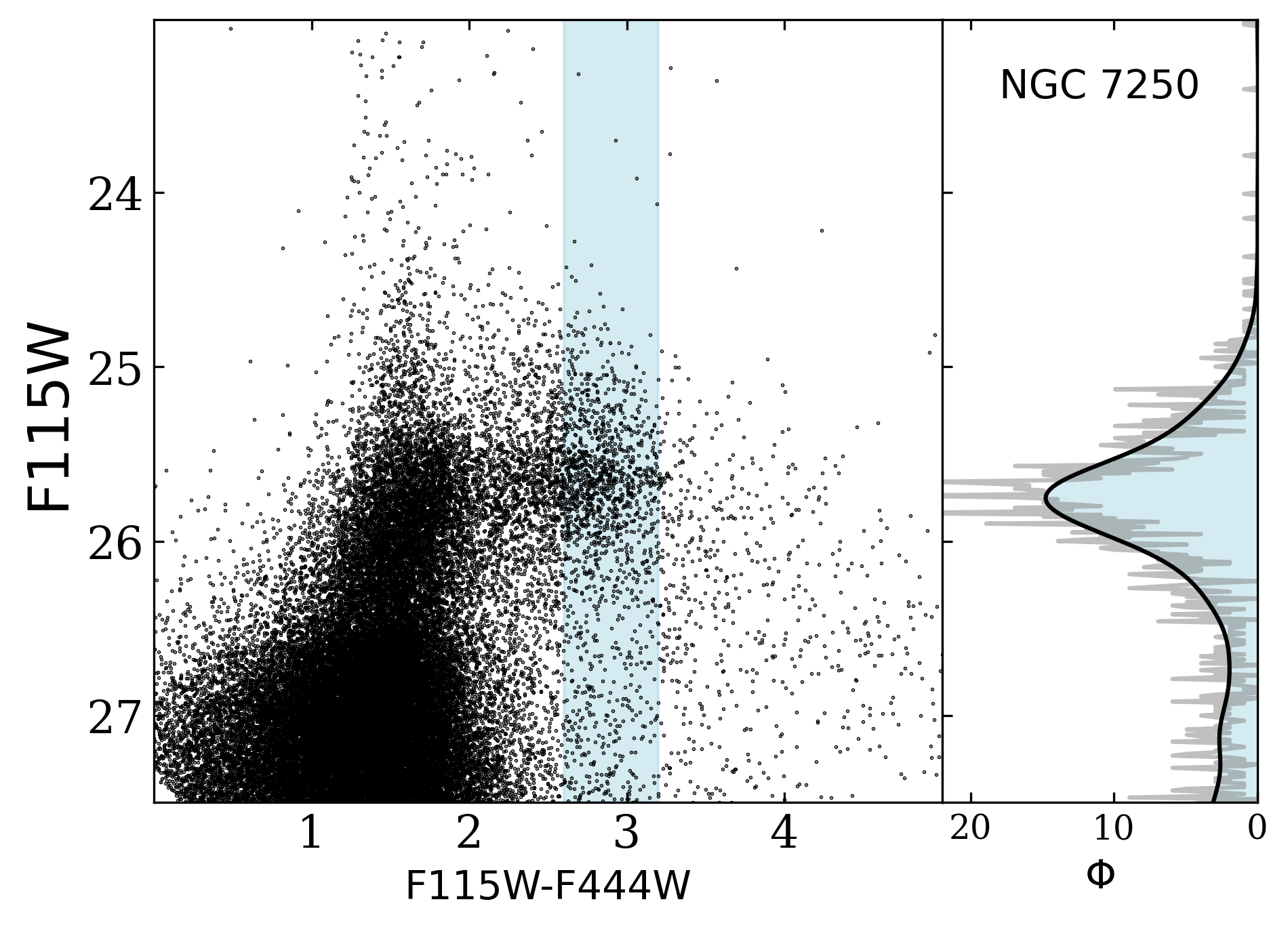}{.5\textwidth}{}}
\gridline {\fig{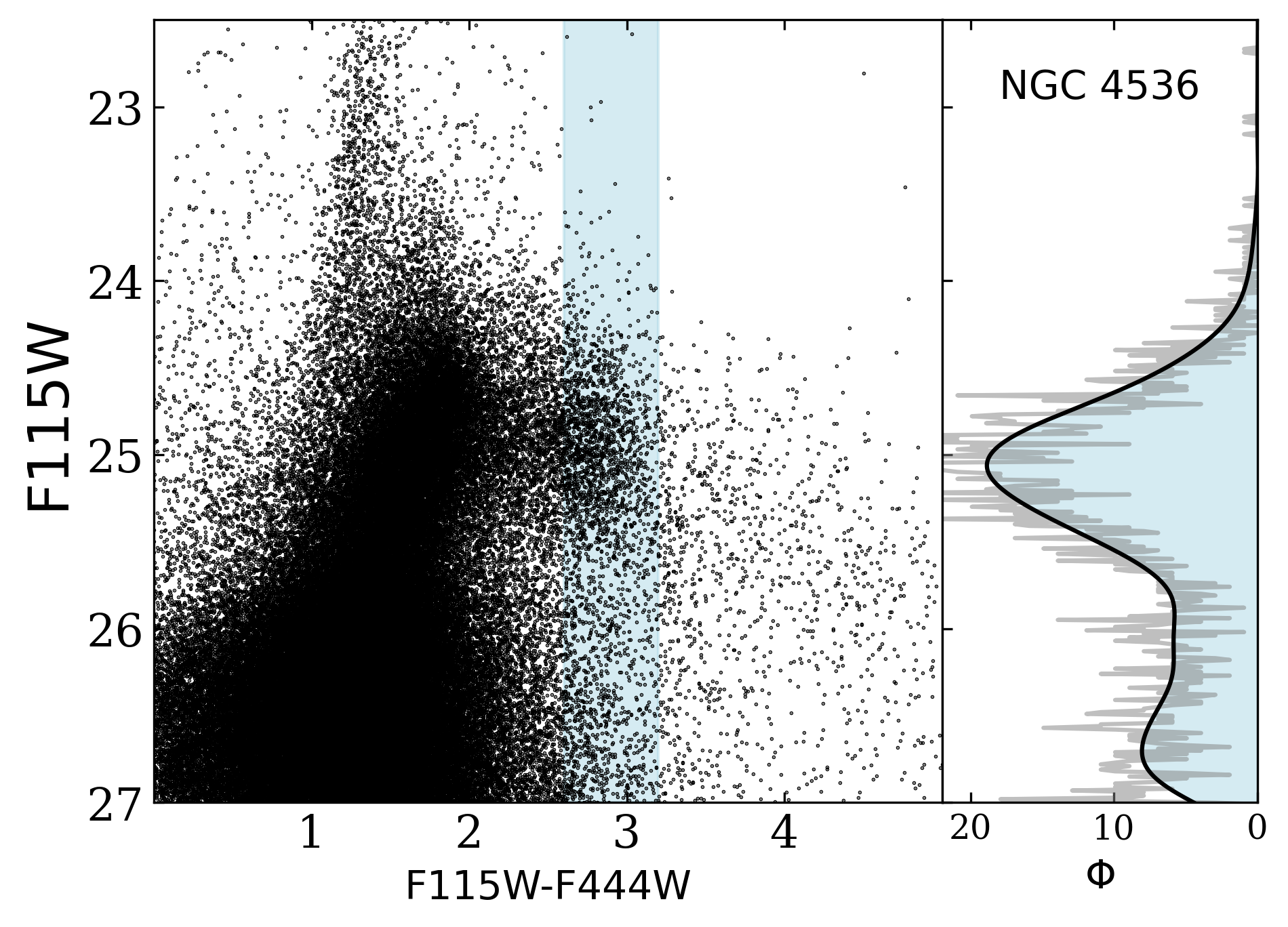}{.5\textwidth}{}
\fig{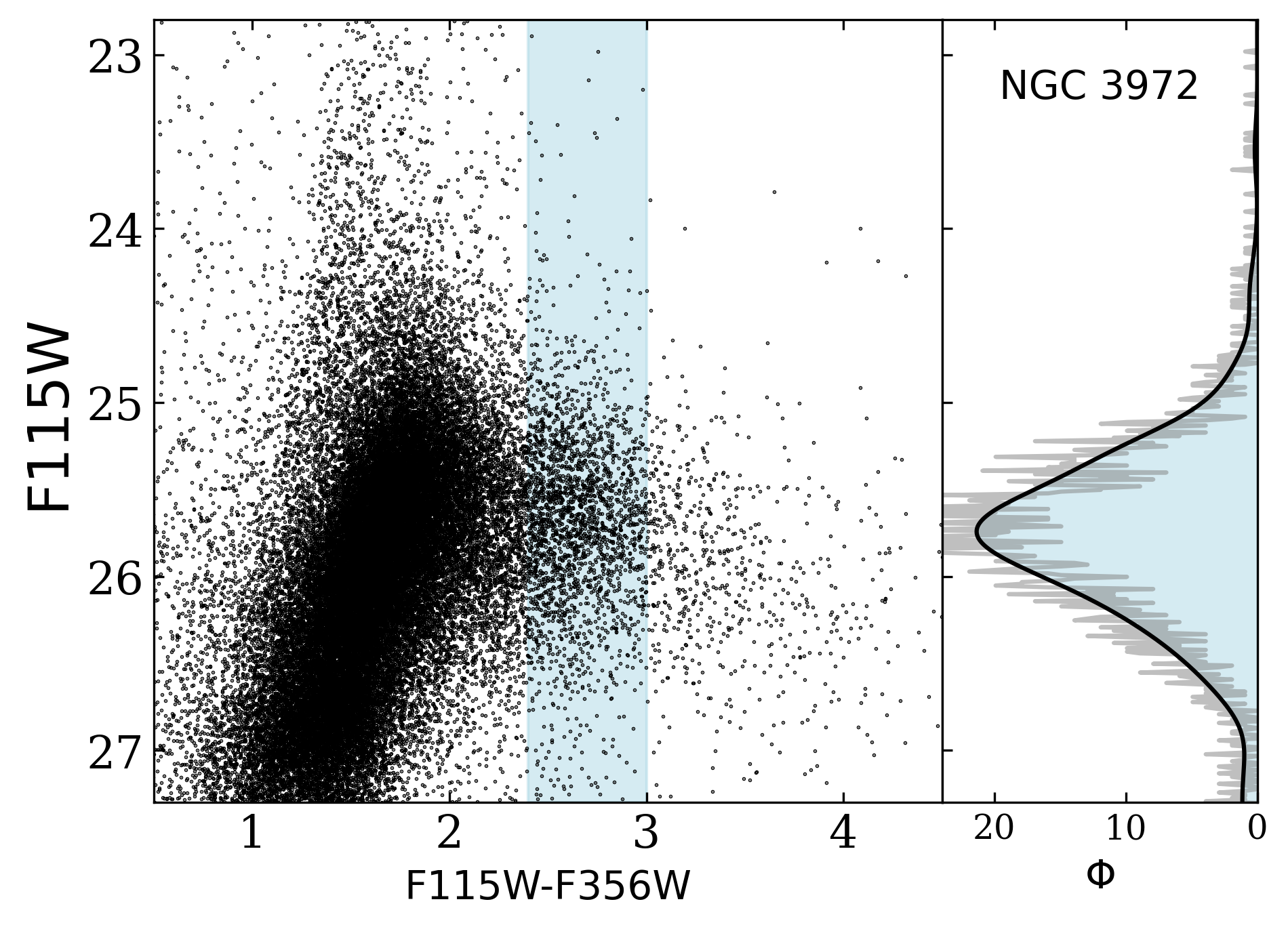}{.5\textwidth}{}}
\caption{
 Color-magnitude diagrams for the outer regions of NGC 7250, NGC 4536, and NGC 3972. The JAGB stars were measured to be within the light-blue shaded regions. In the right-hand panel, the GLOESS-smoothed luminosity functions for the JAGB stars is shown in light blue, over-plotted the 0.01~mag binned luminosity functions in grey. The range of the F115W y-axis is 4.5~mag for all three color magnitude diagrams. 
\label{fig:CMDs}}
\end{figure*}

First, the JAGB stars were identified by their near-infrared color. For NGC~7250 and NGC~4536, JAGB stars were selected as having colors of $2.6<(F115W-F444W)<3.2$~mag. For NGC~3972, JAGB stars were selected as having colors of $2.4<(F115W-F356W)<3.0$~mag. Tests of these color limits are described later in this section.

Then, the  $F115W$ magnitudes of the JAGB stars were finely binned using bin sizes of 0.01~mag. To control for Poisson noise, the binned luminosity function was smoothed using the GLOESS (Gaussian-windowed, LOcally weighted Scatterplot Smoothing) algorithm (introduced by \citealt{2004AJ....128.2239P}), a data-smoothing interpolating technique effective at suppressing false (noise-induced) edges and peaks in luminosity functions. We also describe our tests of varying the smoothing parameters used in the GLOESS algorithm later in this section. The mode of the smoothed luminosity function then marks the JAGB magnitude.

\subsection{Defining the `outer disk'}
\cite{2022ApJ...933..201L} partitioned the photometry of M33 into four concentric regions, showing that the measured mode of the JAGB star luminosity function stabilized to within 0.01~mag in the two outer regions, which covered the outer disk and halo of M33, respectively. The two distance moduli measured in the outer regions agreed with independent TRGB and Leavitt law distance moduli by 2\% \citep{2022ApJ...933..201L}. On the other hand, the mode measured in the inner regions of M33 differed from the mode measured in the outer regions up to $\sim 0.7$~mag. This test first demonstrated that the JAGB magnitude is less accurate when measured in the inner disks of galaxies due to a confluence of crowding, blending, and reddening errors. However, where the `outer disk' lies can be subjective.

\begin{figure*}
\centering
\gridline{\fig{"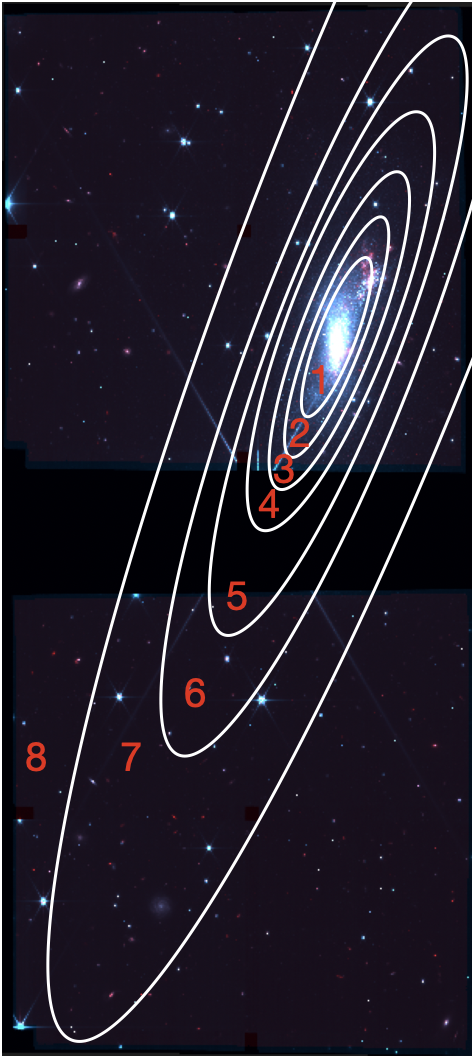"}{0.38\textwidth}{}
\fig{"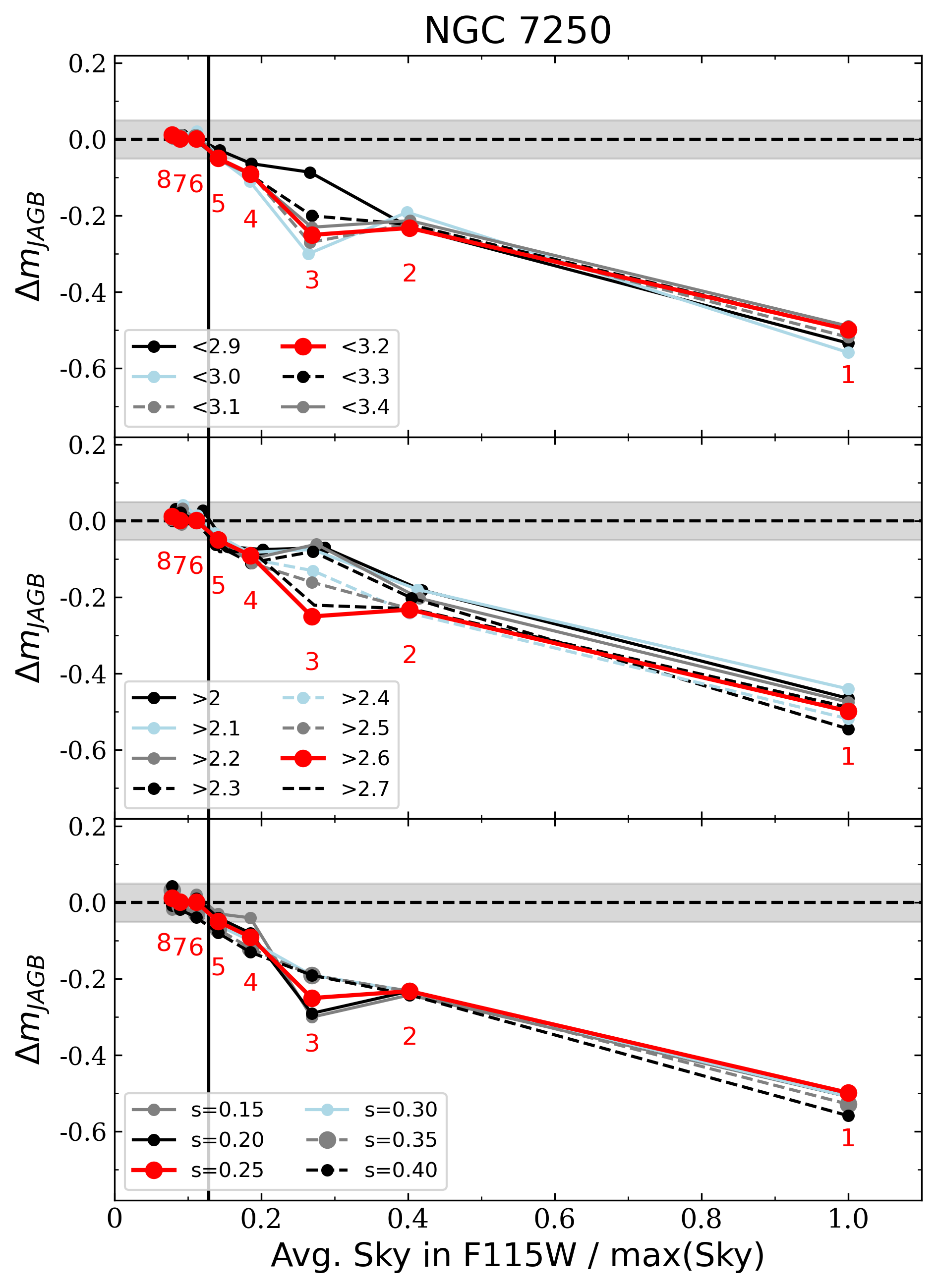"}{0.62\textwidth}{} 
            }
\caption{(Left) 8 regions in NGC 7250, split by semi-major axis. (Right) Change in $\Delta m_{JAGB}$ as a function of the normalized average sky value in that region for the 8 regions. $\Delta m_{JAGB}$  is defined as the change in mode from the fiducial $m_{JAGB}$, which is measured from the outer disk stars between $2.6<(F115W-F444W)<3.2$~mag, using a smoothing parameter for the JAGB star luminosity function of $\sigma_s=0.25$~mag. The top panel shows the effect of this test for different red color cuts while keeping the blue color cut constant, the middle panel shows the effect of this test for different blue color cuts while keeping the red color cut constant, and the bottom panel shows the effect of this test for different smoothing parameters. The black line shows the fiducial surface brightness cutoff used for each galaxy. The grey region in the right plots covers $-0.05$ to 0.05~mag, highlighting the convergence of the JAGB magnitude to within 0.01~mag in Regions 6, 7, and 8. }  
\label{fig:n7250_settledown}
\end{figure*}

\begin{figure*}
\centering
\gridline{\fig{"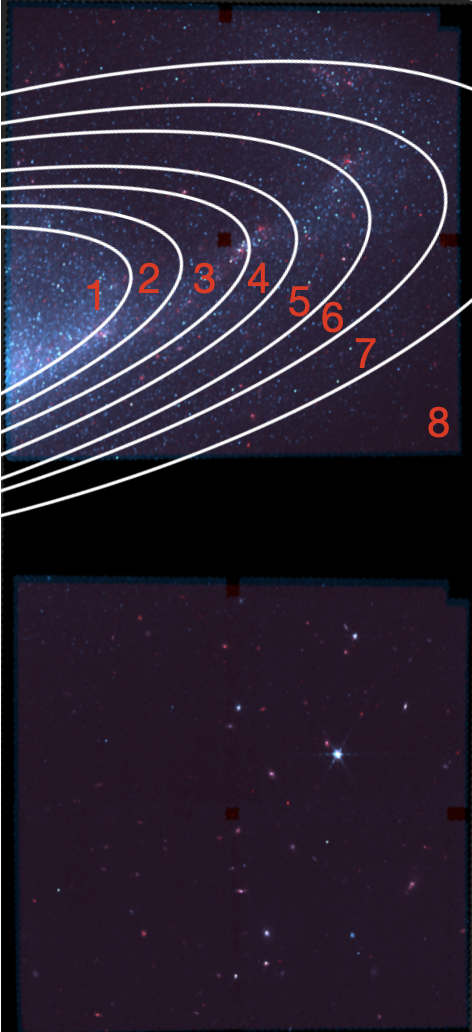"}{0.38\textwidth}{}
\fig{"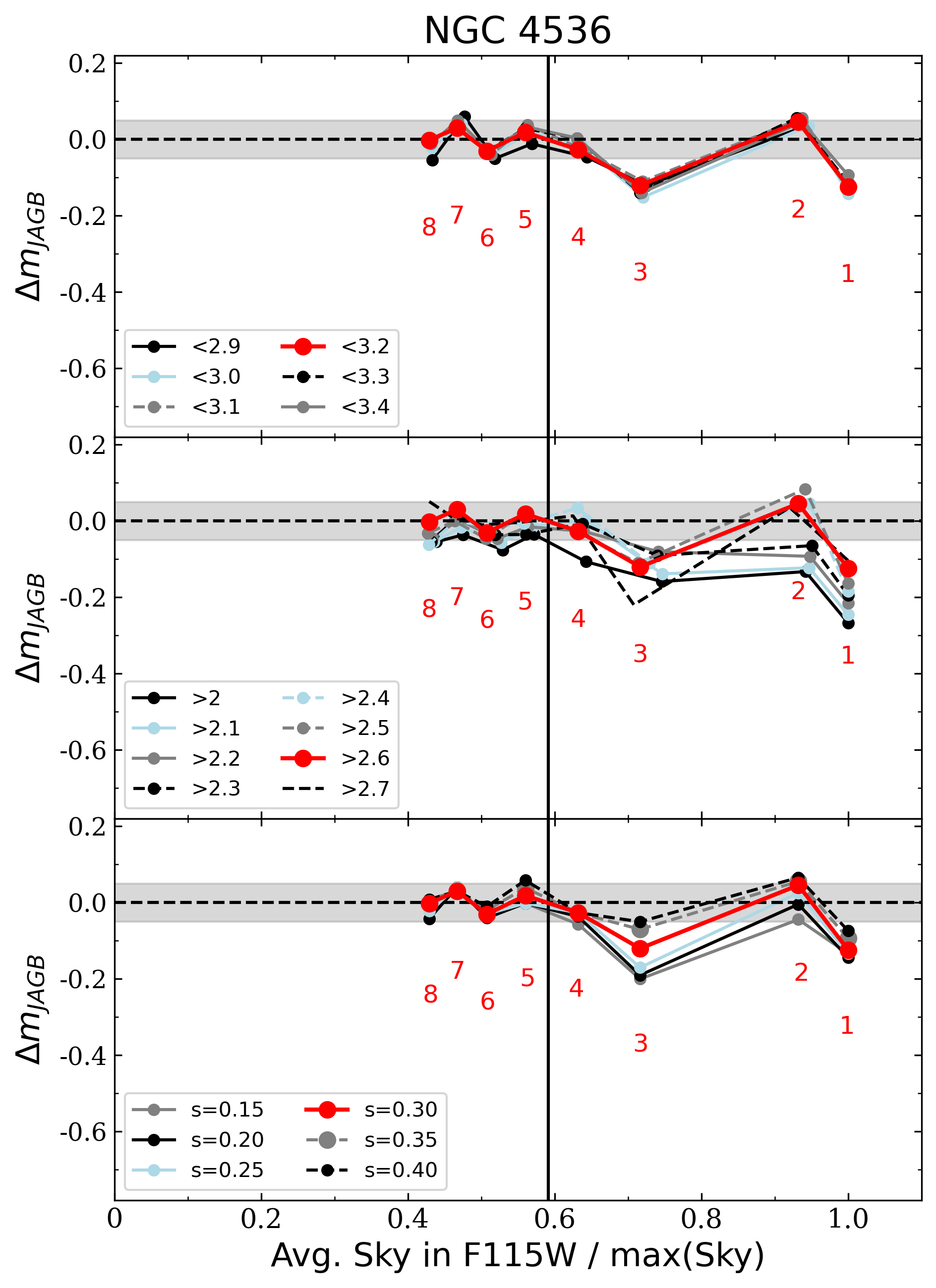"}{0.62\textwidth}{}
            }
\caption{(Left) 8 regions in NGC 4536, split by semi-major axis. (Right) Change in $\Delta m_{JAGB}$ as a function of the normalized average sky value in that region for the 8 regions. $\Delta m_{JAGB}$  is defined as the change in mode from the fiducial $m_{JAGB}$, which is measured for the outer disk stars between $2.6<(F115W-F444W)<3.2$~mag, using a fixed smoothing parameter for the luminosity function of $\sigma_s=0.30$~mag. The top panel shows the effect of this test for different red color cuts while keeping the blue color cut constant, the middle panel shows the effect of this test for different blue color cuts while keeping the red color cut constant, and the bottom panel shows the effect of this test for different smoothing parameters. The black line shows the fiducial surface brightness cutoff used for each galaxy. The grey region in the right plots covers $-0.05$ to 0.05~mag, highlighting the convergence of the JAGB magnitude to within 0.03~mag in regions 5, 6, 7, and 8. }  
\label{fig:n4536_settledown}
\end{figure*}

\begin{figure*}
\centering
\gridline{\fig{"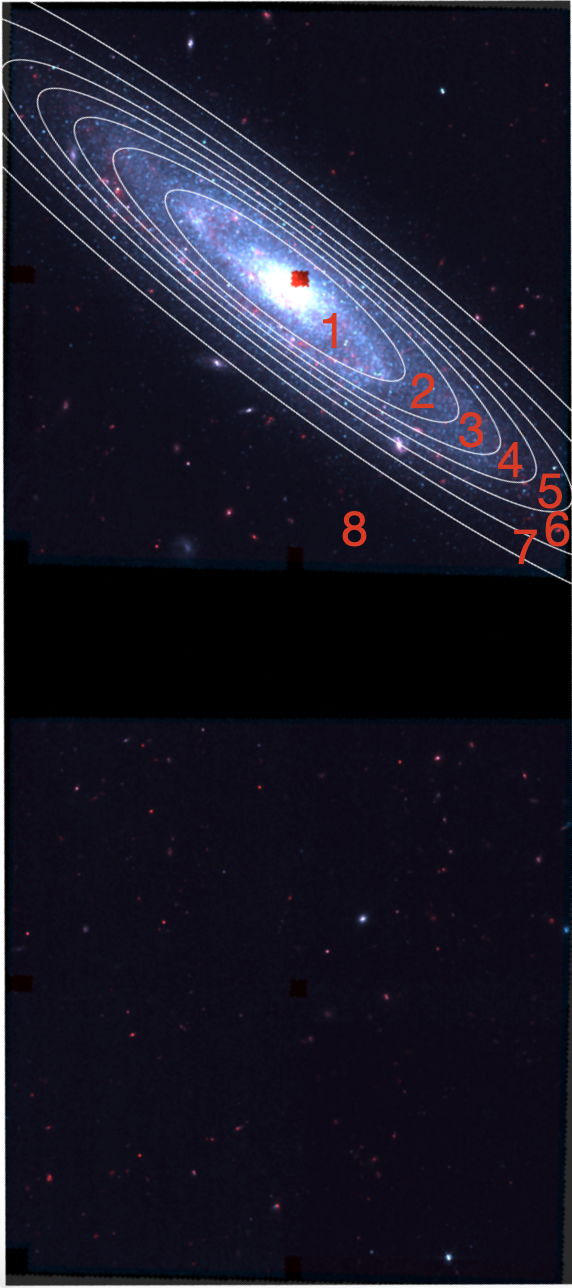"}{0.38\textwidth}{}
\fig{"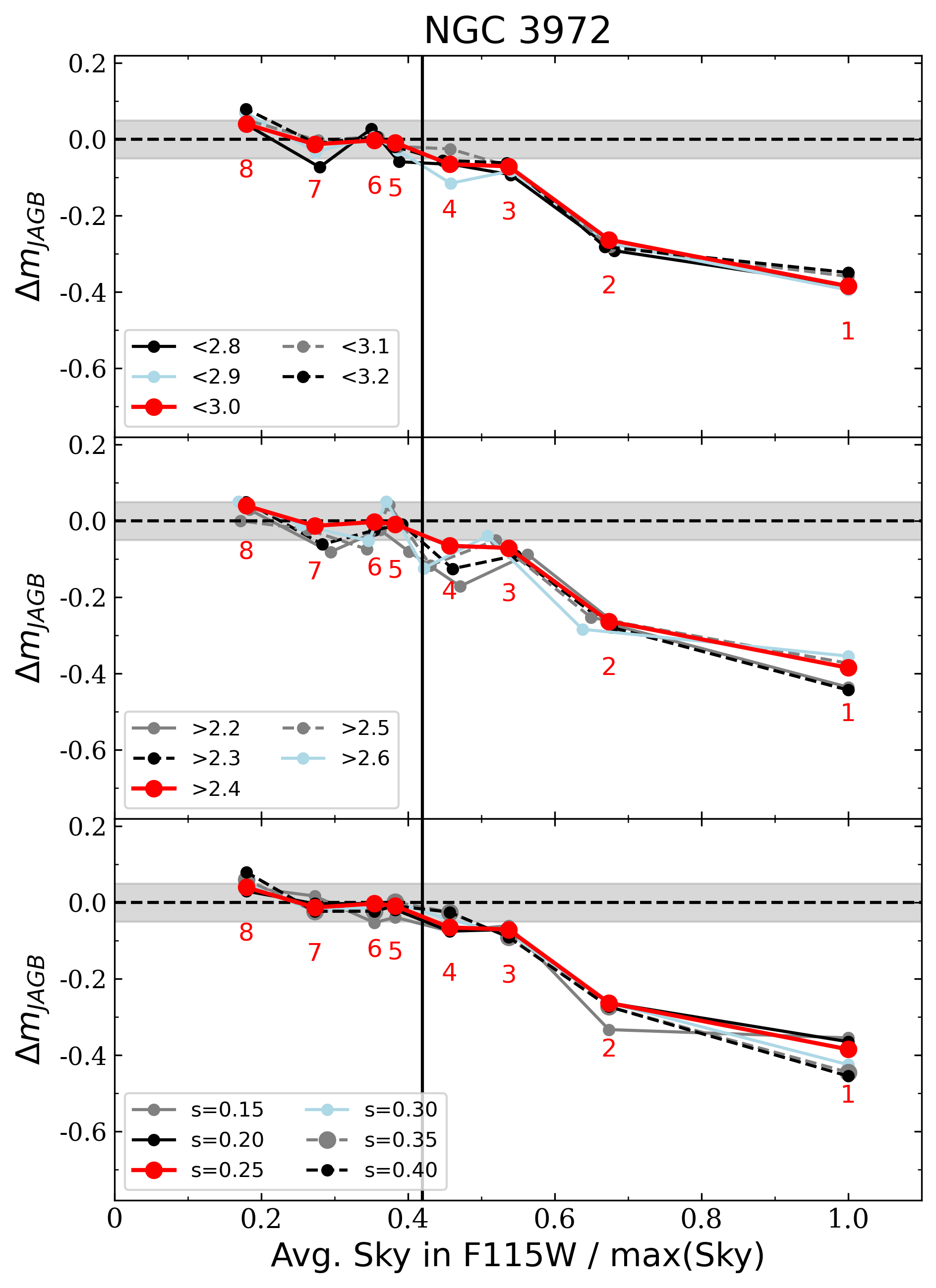"}{0.62\textwidth}{}
            }
\caption{(Left) 8 regions in NGC 3972, split by semi-major axis. (Right) Change in $\Delta m_{JAGB}$ as a function of the normalized average sky value in that region for the 8 regions. $\Delta m_{JAGB}$  is defined as the change in mode from the fiducial $m_{JAGB}$, which is measured for the outer disk stars between $2.4<(F115W-F444W)<3.0$~mag, using a smoothing parameter for the luminosity function of $\sigma_s=0.25$~mag. The top panel shows the effect of this test for different red color cuts while keeping the blue color cut constant, the middle panel shows the effect of this test for different blue color cuts while keeping the red color cut constant, and the bottom panel shows the effect of this test for different smoothing parameters. The black line shows the fiducial surface brightness cutoff used for each galaxy. The grey region in the right plots covers $-0.05$ to 0.05~mag, highlighting the convergence of the JAGB magnitude to within 0.04~mag in Regions 5, 6, 7, and 8. }  
\label{fig:n3972_settledown}
\end{figure*}

We introduce a novel method for identifying the outer disk of galaxies for the purpose of measuring JAGB method distances, where reddening, crowding, and blending effects taper off. We separated the photometry of a given galaxy into eight concentric regions split by semi-major axis (SMA), and measured the change in JAGB magnitude from region to region. To compute the SMA, we obtained the galaxy's center coordinates from NED, and its inclination angle and position angle from HyperLeda.
The boundaries of the regions were chosen so that each region had an equal number of JAGB stars. These boundaries are shown in the left panels of Figures \ref{fig:n7250_settledown}, \ref{fig:n4536_settledown}, and \ref{fig:n3972_settledown}, for NGC 7250, NGC 4536, and NGC 3972, respectively, overlaid onto color mosaics of our NIRCam imaging.

To measure relative changes in mode throughout the different regions, we computed $\Delta m_{JAGB}$, which was defined as the change in mode from the fiducial mode. In NGC 7250, the fiducial mode was measured using JAGB stars with colors between $2.6<(F115W-F444W)<3.2$~mag and an LF smoothed with a smoothing parameter of $\sigma_s=0.25$~mag.
In NGC 4536, the fiducial mode was measured using JAGB stars with colors between $2.6<(F115W-F444W)<3.2$~mag and an LF smoothed with a smoothing parameter of $\sigma_s=0.30$~mag. For NGC 3972, the fiducial mode was measured using JAGB stars with colors between $2.4<(F115W-F444W)<3.0$~mag and an LF smoothed with a smoothing parameter of $\sigma_s=0.25$~mag. 
Regarding the spatial selection, the fiducial JAGB modal magnitude was computed from the final selected `outer disk' region (i.e., the merged outer regions to the left of the black line in Figure \ref{fig:appendix}.). We now describe our process for selecting this outer disk region below.
The color magnitude diagrams with these chosen parameters are shown in Figure \ref{fig:CMDs}. 

We then compared $\Delta m_{JAGB}$ to the \textit{relative} average surface brightness in each radial region.\footnote{Using a galaxy's surface brightness profile to determine the optimal halo location has also been used by the CCHP in measuring the TRGB (e.g., \citealt{2021ApJ...906..125J}).}
In the right panels of Figures \ref{fig:n7250_settledown}, \ref{fig:n4536_settledown}, and \ref{fig:n3972_settledown}, we show the change in $\Delta m_{JAGB}$ as a function of the average \texttt{sky\_{F115W}}\footnote{Labeled the \texttt{total sky value} in the DOLPHOT catalog.}  value in each radial bin, divided by the maximum \texttt{sky\_{F115W}} value for the whole galaxy (all concentric regions).
The \texttt{sky\_{F115W}} parameter was derived from DOLPHOT for every star, and is a relative measure of the sky background at the star's position in a given image \citep{2016ascl.soft08013D}. 
In Appendix \ref{sec:appendix}, we show plots of this \texttt{sky\_{F115W}} parameter as a function of the semi-major axis, with  corresponding power-law fits. We emphasize the \texttt{sky\_{F115W}} parameter is not an exact measure of surface brightness but is a simple parameter returned from DOLPHOT that can be used to track changes in $\Delta m_{JAGB}$. We also highlight that the choice of region separation (we used SMA here) is irrelevant to the final result, as long as $\Delta m_{JAGB}$ converges in the outermost regions.

The red points in Figures \ref{fig:n7250_settledown}, \ref{fig:n4536_settledown}, and \ref{fig:n3972_settledown} show the change in $\Delta m_{JAGB}$ as a function of the \texttt{sky\_{F115W}} parameter in each galaxy. The three panels show the effects of using different color cuts and smoothing parameters. In all cases, the JAGB magnitude leveled out in the outermost regions of the galaxy; therefore, the choice of color cuts and smoothing parameter did not significantly affect the final result. 
For example in NGC 7250 and NGC 4536, where we used a color cut of $2.6<(F115W-F444W)<3.2$~mag to measure the fiducial JAGB magnitude, we varied the blue color cut through \{2.0, 2.1, 2.2, 2.3, 2.4, 2.5, 2.6, 2.7\} and the red color through \{2.9, 3.0, 3.1, 3.2, 3.3, 3.4\}. For NGC 3972 where we used a color cut of $2.4<(F115W-F356W)<3.0$~mag to measure the fiducial JAGB magnitude, we varied the blue color cut through \{2.2, 2.3, 2.4, 2.5, 2.6\} and the red color through \{2.8, 2.9, 3.0, 3.1, 3.2\}.
We also tested how different smoothing parameters $\sigma_s$ in the GLOESS smoothing algorithm affected the mode. We varied $\sigma_s$ from 0.15 to 0.40~mag in steps of 0.05~mag and quantified the change in the mode. 

In all three galaxies, the mode $m_{JAGB}$ stabilized in the outer regions of the galaxy.
In NGC 7250, each region had 335 JAGB stars. The mode $m_{JAGB}$ stabilized to within 0.01 in the three outermost regions.
In NGC 4536, each region had 461 JAGB stars. The mode $m_{JAGB}$ stabilized to within 0.03 in the five outermost regions.
In NGC 3972, each region had 491 JAGB stars. The mode $m_{JAGB}$ stabilized to within 0.04 in the four outermost regions. 
 Therefore, we selected the outer bin of the innermost stable region to be the radial cutoff for that galaxy. In NGC 7250, this was $9.6$~kpc; for NGC 4536, this was $SMA=10.3$~kpc; for NGC 4972, this was $SMA=7.6$~kpc.
We recommend this procedure be used for measuring the JAGB magnitude for all future studies utilizing the JAGB method; the inner disk regions of galaxies can therefore be cut out in a procedural way, instead of visually.

\subsection{Color-magnitude diagrams}

The color-magnitude diagrams and smoothed luminosity functions for all three galaxies after performing the inner disk spatial cut are shown in Figure \ref{fig:CMDs}. The scatter on the mode $m_{JAGB}$ of the JAGB stars between $m_{JAGB}\pm0.75$~mag was measured to be $\sigma=0.32$~mag, $\sigma=0.34$~mag, and $\sigma=0.35$~mag for NGC 7250, NGC 4536, and NGC 3972, respectively. To measure the scatter on the mode, we used the following formula:

\begin{equation}
    \sqrt {\frac{\Sigma{(m_i -m_{JAGB})^2}}{N_{JAGB}} },
\end{equation}

\noindent
where $m_{i}$ are the F115W magnitudes of the JAGB stars and $N_{JAGB}$ is the total number of JAGB stars.

These measured scatters from JWST are similar to the scatter measured from ground-based telescopes for the JAGB stars in the LMC, a galaxy 400 times closer than NGC 7250. 
\cite{2001ApJ...548..712W}, the first study to use JAGB stars as standard candles in the near infrared, measured a scatter of $\sigma=0.33$~mag in their LMC sample using 2MASS data.

\subsection{Artificial star tests}
\begin{figure*}
\centering
\includegraphics[width=\textwidth]{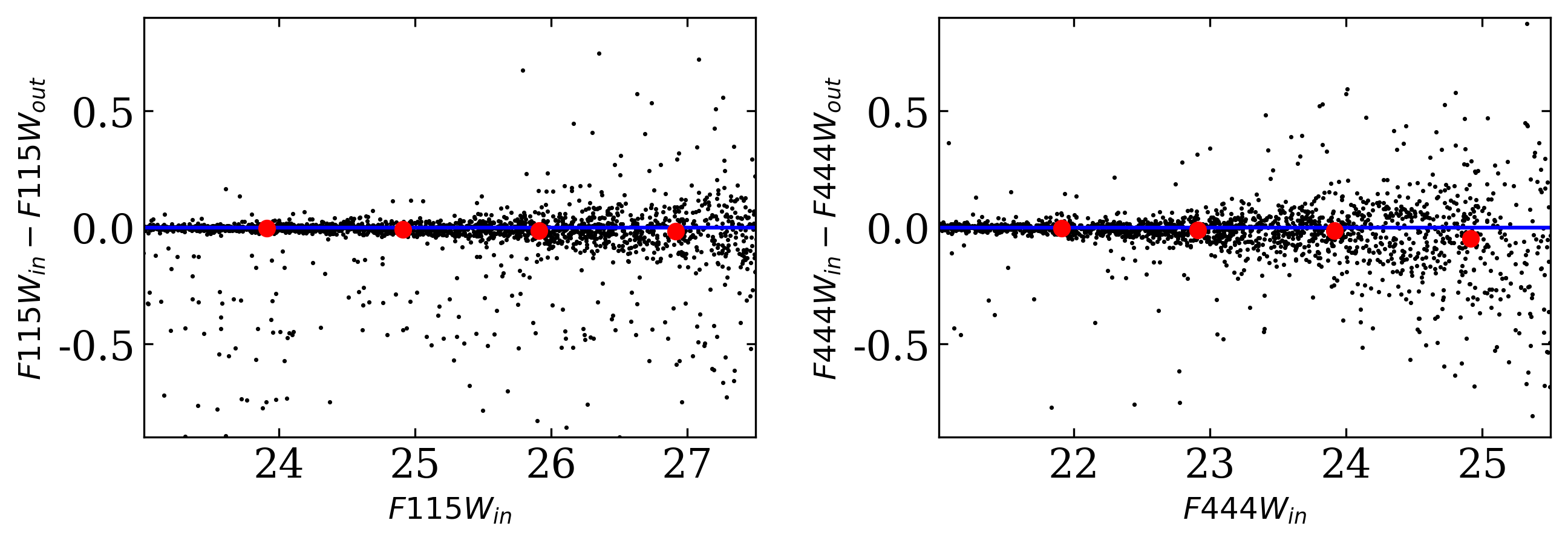}
\caption{Difference between the input and output magnitudes from the DOLPHOT artificial star tests for the outer region of NGC 7250 ($d>9.6$~kpc). The red dots represent the median offset for 1~mag bins. The same random offset between $-0.1$ and $0.1$~mag discussed in Section \ref{sec:data} has been applied to both magnitude axes.
} 
\label{fig:fakestars}
\end{figure*}

Artificial stars were computed with DOLPHOT for the first galaxy in our program, NGC 7250, to assess the robustness of the JAGB star photometry. In Figure \ref{fig:fakestars}, we show results for the $\sim4200$ injected stars in the JAGB color range and having a $SMA>9.6$~kpc. The range of the x-axis in Figure \ref{fig:fakestars} corresponds to the same range of the y-axis in Figure \ref{fig:CMDs} for NGC 7250. At the average magnitude of the JAGB stars, the median offset was measured to be $F115W_{in}-F115W_{out}=-0.01$~mag, demonstrating that crowding is a negligible source of systematic error for the JAGB stars in the outer regions of NGC 7250. 

\section{Conclusion}
This paper demonstrates JWST's impressive capabilities for utilizing JAGB stars as standard candles. Our results exemplify we now have the potential to study JAGB stars in galaxies at distances $\gtrsim20$~Mpc away with JWST, with comparable or greater resolution than galaxies $50$~kpc away with ground-based telescopes. 

The high sensitivity and resolution of JWST allows us to resolve JAGB and TRGB stars in the outer regions of galaxies, at much farther distances than what was previously possible with HST.
For example, all of the Cepheid variable stars in NGC 7250 have been discovered in the inner region (\citealt{2016ApJ...826...56R, 2022ApJ...934L...7R}, Owens et al., submitted), i.e., within the white ellipse in Figure \ref{fig:montage}, where the JAGB measurement was excluded in this paper.
Thus, a significant advantage of the JAGB and TRGB methods to the Cepheid P-L relation is that carbon stars can be found in the significantly less crowded halos of galaxies, whereas Cepheids are mainly found in the more crowded star-forming disks.

A future calibration of the JAGB stars absolute magnitude in the  $F115W$ filter will soon be possible via imaging of the JAGB stars in the water mega-maser host galaxy NGC 4258. These images are now scheduled for early 2024 and will also provide a zero point for the TRGB and Leavitt law. This calibration will allow us to determine the distance to NGC\,7250, NGC 4536, and NGC 3972 via the JAGB stars, as well as the rest of the SN Ia host galaxies in our observing program, allowing us to independently measure $H_o$ via the JAGB method.

\begin{acknowledgments}

Suggestions and comments from Saurabh Jha, Adam Riess, and Dan Scolnic at the 2023 MIAPbP extragalactic distance scale workshop, regarding the selection of the location for the JAGB measurement, are gratefully acknowledged. We thank Dan Weisz and Andy Dolphin for their insight and advice on the use of the latest version of DOLPHOT tailored to JWST
imaging data, and Jane Rigby for updates on the calibration of the NIRCam filters. Finally, we thank the anonymous referee for their constructive and helpful suggestions that improved this work.

AJL was supported by the Future Investigators in NASA Earth and Space Science
and Technology (FINESST) award number 80NSSC22K1602 during the completion of
this work.
AJL thanks the LSSTC Data Science Fellowship Program, which is funded by LSSTC, NSF Cybertraining Grant \#1829740, the Brinson Foundation, and the Moore Foundation; her participation in the program has benefited this work. 
AJL was supported by the Munich Institute for Astro-, Particle and BioPhysics (MIAPbP) which is funded by the Deutsche Forschungsgemeinschaft (DFG, German Research Foundation) under Germany´s Excellence Strategy – EXC-2094 – 390783311.
We also thank the University of Chicago and the Observatories of the Carnegie Institution for their support of our long-term research into the calibration and determination of the expansion rate of the universe. 

This work is based on observations made with the NASA/ESA/CSA JWST. The data were obtained from the Mikulski Archive for Space Telescopes at the Space Telescope Science Institute, which is operated by the Association of Universities for Research in Astronomy, Inc., under NASA contract NAS 5-03127 for JWST. These observations are associated with program 1995.
This research has made use of NASA's Astrophysics Data System Bibliographic Services.

All the {\it JWST} data used in this paper can be found in MAST: \dataset[10.17909/ffv8-5279]{http://dx.doi.org/10.17909/ffv8-5279}.

\end{acknowledgments}

\software{NumPy \citep{2020Natur.585..357H}, Matplotlib \citep{2007CSE.....9...90H}, Pandas \citep{pandas}}

\facilities{JWST (NIRCam)}

\appendix
\restartappendixnumbering

\section{Sky count profiles}\label{sec:appendix}
\begin{figure*}
\gridline{\fig{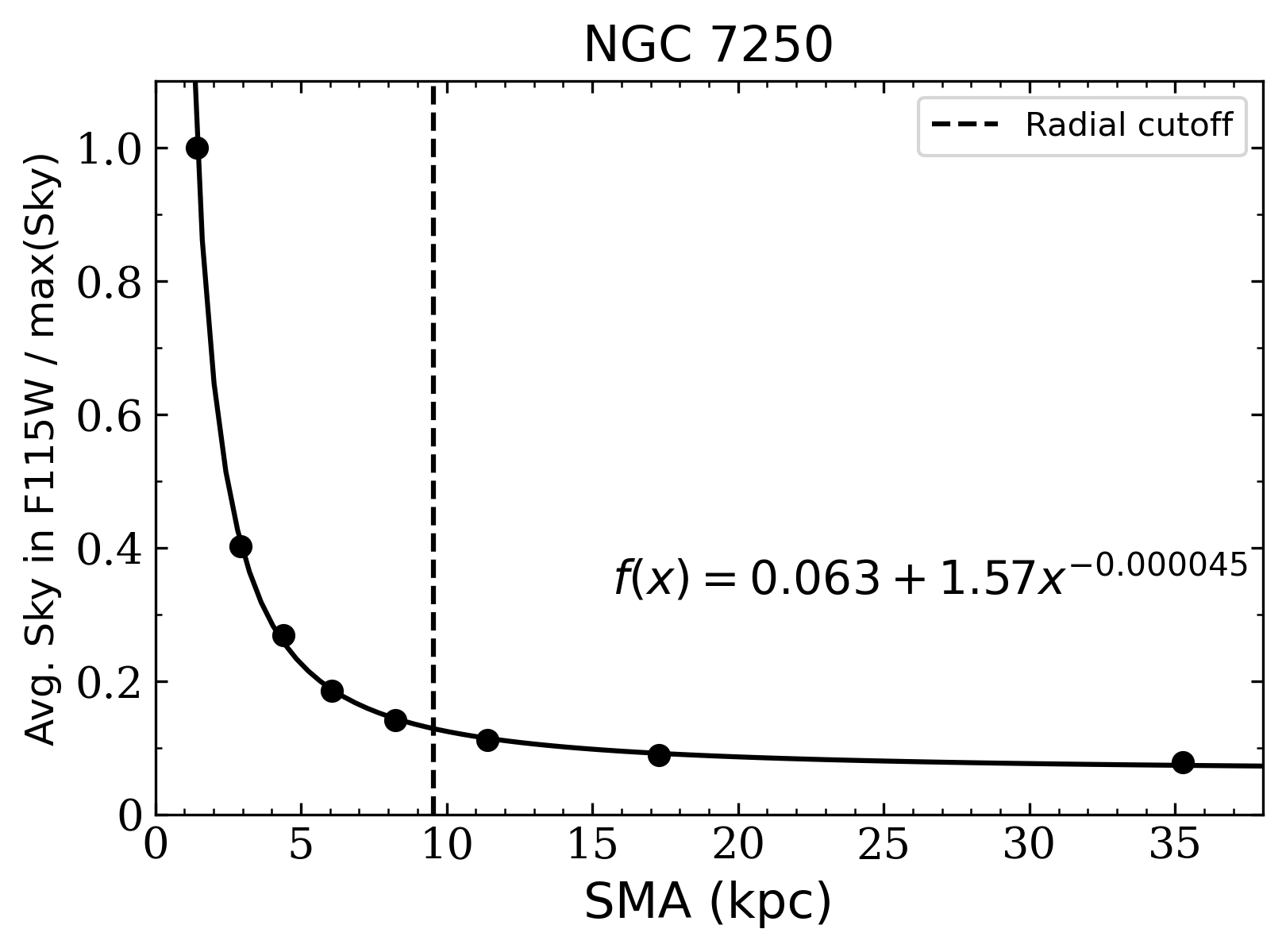}{.5\textwidth}{}}
\gridline{\fig{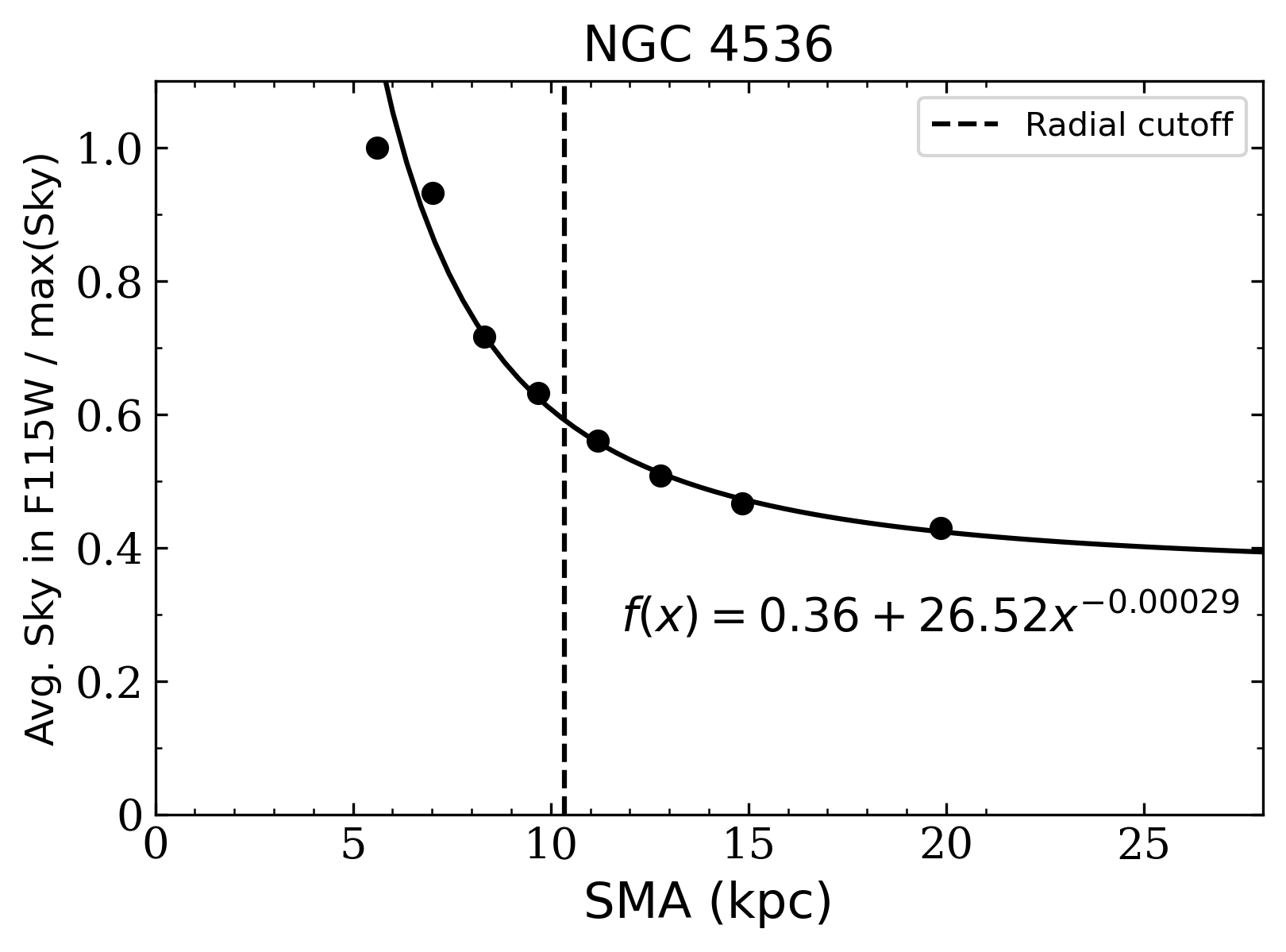}{.5\textwidth}{}
\fig{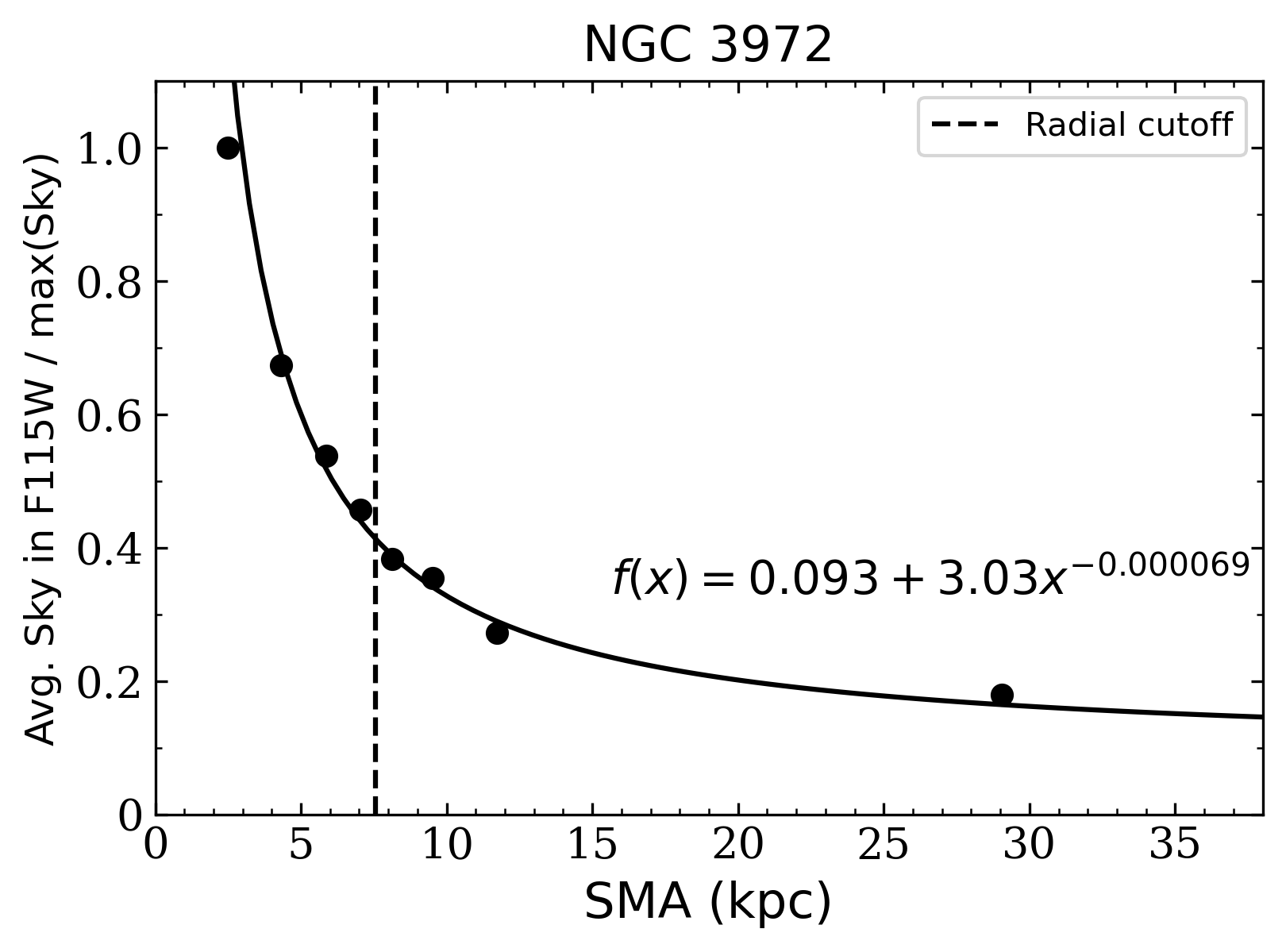}{.5\textwidth}{}}
\caption{Average normalized \texttt{sky\_{F115W}} parameter returned from DOLPHOT as a function of SMA for NGC 7250, NGC 4536, and NGC 3972, for the 8 defined regions split by SMA. The average \texttt{sky\_{F115W}} and SMA for each region is shown by the black points. Each region has the same number of JAGB stars. Power law functions were fit to the points and shown here as black lines. The analytical form of the power law function is shown on each plot.} The dotted black line is the fiducial radial cut that was used for each galaxy. The corresponding \texttt{sky\_{F115W}} returned from the power-law fit for the chosen radial cutoff is shown in Figures \ref{fig:n7250_settledown}, \ref{fig:n4536_settledown}, and \ref{fig:n3972_settledown}.
\label{fig:appendix}
\end{figure*}

In this section, we show plots of the average \texttt{sky\_F115W} parameter in each of the 8 regions as a function of semi-major axis for the three galaxies. Power-law functions were fit to the data points, and are shown as the black line in Figure \ref{fig:appendix}.

\begin{thebibliography}

\bibitem[Aiola et al.(2020)]{2020JCAP...12..047A} Aiola, S., Calabrese, E., Maurin, L., et al.\ 2020, \jcap, 2020, 047

\bibitem[Boyer et al.(2022)]{2022RNAAS...6..191B} Boyer, M.~L., Anderson, J., Gennaro, M., et al.\ 2022, Research Notes of the American Astronomical Society, 6, 191

\bibitem[Di Valentino et al.(2021)]{2021arXiv210301183D} Di Valentino, E., Mena, O., Pan, S., et al.\ 2021, Classical and Quantum Gravity, 38, 153001.

\bibitem[Dolphin(2000)]{2000PASP..112.1383D} Dolphin, A.~E.\ 2000, \pasp, 112, 1383

\bibitem[Dolphin(2016)]{2016ascl.soft08013D} Dolphin, A.\ 2016, Astrophysics Source Code Library. ascl:1608.013

\bibitem[Freedman(2021)]{2021ApJ...919...16F} Freedman, W.~L.\ 2021, \apj, 919, 16.

\bibitem[Freedman et al.(2021)]{2021jwst.prop.1995F} Freedman, W.~L., Madore, B.~F., Hoyt, T., et al.\ 2021, JWST Proposal. Cycle 1, 1995

\bibitem[Freedman \& Madore(2020)]{2020ApJ...899...67F} Freedman, W.~L. \& Madore, B.~F.\ 2020, \apj, 899, 67


\bibitem[Gardner et al.(2023)]{2023arXiv230404869G} Gardner, J.~P., Mather, J.~C., Abbott, R., et al.\ 2023, \pasp, 135, 068001

\bibitem[Gordon et al.(2022)]{2022AJ....163..267G} Gordon, K.~D., Bohlin, R., Sloan, G.~C., et al.\ 2022, \aj, 163, 267

\bibitem[Habing \& Olofsson(2003)]{2003agbs.conf.....H} Habing, H.~J. \& Olofsson, H.\ 2003, Asymptotic giant branch stars (New York: Springer)

\bibitem[Hatt et al.(2018)]{2018ApJ...861..104H} Hatt, D., Freedman, W.~L., Madore, B.~F., et al.\ 2018, \apj, 861, 104


\bibitem[Hunter(2007)]{2007CSE.....9...90H} Hunter, J.~D.\ 2007, Computing in Science and Engineering, 9, 90

\bibitem[Iben(1973)]{1973ApJ...185..209I} Iben, I.\ 1973, \apj, 185, 209

\bibitem[Iben \& Renzini(1983)]{1983ARA&A..21..271I} Iben, I. \& Renzini, A.\ 1983, \araa, 21, 271

\bibitem[Jang et al.(2021)]{2021ApJ...906..125J} Jang, I.~S., Hoyt, T.~J., Beaton, R.~L., et al.\ 2021, \apj, 906, 125

\bibitem[Jang \& Lee(2015)]{2015ApJ...807..133J} Jang, I.~S. \& Lee, M.~G.\ 2015, \apj, 807, 133

 \bibitem[Lee et al.(2021)]{2020arXiv201204536L} Lee, A.~J., Freedman, W.~L., Madore, B.~F., et al.\ 2021, \apj, 907, 112 


\bibitem[Lee et al.(2022)]{2022ApJ...933..201L} Lee, A.~J., Rousseau-Nepton, L., Freedman, W.~L., et al.\ 2022, \apj, 933, 201



\bibitem[Madore \& Freedman(2020)]{2020ApJ...899...66M} Madore, B.~F. \& Freedman, W.~L.\ 2020, \apj, 899, 66

\bibitem[Madore et al.(2022)]{2022ApJ...926..153M} Madore, B.~F., Freedman, W.~L., \& Lee, A.~J.\ 2022, \apj, 926, 153


\bibitem[Marigo et al.(2008)]{2008A&A...482..883M} Marigo, P., Girardi, L., Bressan, A., et al.\ 2008, \aap, 482, 883

\bibitem[McKinney(2010)]{pandas} McKinney, W.\ 2010, Proceedings of the 9th Python in Science Conference, 51-56

\bibitem[Nikolaev \& Weinberg(2000)]{2000ApJ...542..804N} Nikolaev, S. \& Weinberg, M.~D.\ 2000, \apj, 542, 804


\bibitem[Parada et al.(2021)]{2021MNRAS.501..933P} Parada, J., Heyl, J., Richer, H., et al.\ 2021, \mnras, 501, 933

\bibitem[Parada et al.(2023)]{2023MNRAS.522..195P} Parada, J., Heyl, J., Richer, H., et al.\ 2023, \mnras, 522, 195

\bibitem[Persson et al.(2004)]{2004AJ....128.2239P} Persson, S.~E., Madore, B.~F., Krzemi{\'n}ski, W., et al.\ 2004, \aj, 128, 2239

\bibitem[Planck Collaboration et al.(2020)]{2020A&A...641A...6P} Planck Collaboration, Aghanim, N., Akrami, Y., et al.\ 2020, \aap, 641, A6 

\bibitem[Rieke et al.(2023)]{2023PASP..135b8001R} Rieke, M.~J., Kelly, D.~M., Misselt, K., et al.\ 2023, \pasp, 135, 028001

\bibitem[Riess et al.(2016)]{2016ApJ...826...56R} Riess, A.~G., Macri, L.~M., Hoffmann, S.~L., et al.\ 2016, \apj, 826, 56

\bibitem[Riess et al.(2022)]{2022ApJ...934L...7R} Riess, A.~G., Yuan, W., Macri, L.~M., et al.\ 2022, \apjl, 934, L7

\bibitem[Rigby et al.(2023)]{2023PASP..135d8001R} Rigby, J., Perrin, M., McElwain, M., et al.\ 2023, \pasp, 135, 048001

\bibitem[Ripoche et al.(2020)]{2020MNRAS.495.2858R} Ripoche, P., Heyl, J., Parada, J., et al.\ 2020, \mnras, 495, 2858 





\bibitem[Harris et al.(2020)]{2020Natur.585..357H} Harris, C.~R., Millman, K.~J., van der Walt, S.~J., et al.\ 2020, \nat, 585, 357

\bibitem[Weinberg \& Nikolaev(2001)]{2001ApJ...548..712W} Weinberg, M.~D. \& Nikolaev, S.\ 2001, \apj, 548, 712

\bibitem[Weisz et al.(2023)]{2023arXiv230104659W} Weisz, D.~R., McQuinn, K.~B.~W., Savino, A., et al.\ 2023, \apjs, 268, 15


\bibitem[Zgirski et al.(2021)]{2021arXiv210502120Z} Zgirski, B., Pietrzy{\'n}ski, G., Gieren, W., et al.\ 2021, \apj, 916, 19



\end{thebibliography}
\end{document}